# A Discussion on Stabilization of Frequency Control for Power Systems

Binh-Minh Nguyen, Ngoc Tran-Huynh, Michihiro Kawanishi, and Tatsuo Narikiyo
Control System Laboratory, Toyota Technological Institute, Aichi, Japan

*Abstract* - How to practically maintain the frequency stability of large-scale power systems by a decentralized way is a simple but non-trivial question. In other words, is it possible to design any local controller without understanding the other controlled areas and the details of network structure? With respect to the special properties of physical interaction between the local areas, this paper suggests two existing theories for tackling this issue. Firstly, passivity theory is shown to be a candidate for frequency control problem using swing equation. Based on the passivity of swing dynamics, it is possible to guarantee the system stability by designing for each local area a passive controller. We further extend the passivity approach to the hierarchically decentralized control system with unknown communication delay. Secondly, we are motivated by the frequency control problem using area-control-error. Each local controller is designed such that each local subsystem follows a nominal model $H_n(s)$ of volume $\xi$. Utilizing generalized frequency variable (GFV) theory, we present a triad of conditions that sufficiently guarantee the system stability. The conditions can be tested conveniently by a limited set of inequalities established from the GFV $\Phi_n(s) = 1/H_n(s)$ and the eigenvalues of the physical interaction matrix. The effectiveness, limitation, and challenge of two theories are discussed by design examples with numerical simulations.

*Keywords* - decentralized stabilization, generalized frequency variable, load frequency control, passivity.

## 1. Introduction

Maintaining a steady frequency is one of the fundamental issues in the field of power system control. It is well-known that if the frequency deviation exceeds a certain limit, generation and utilization equipment might cease to operate properly [1]. Due to the increasing in size of power systems and the introduction of renewable energies, researches on centralized frequency control [2] and single-area-LFC [3], [4], [5], [6] are no longer of interest. As listed in TABLE I, various decentralized frequency control methods have been proposed [7] ~ [20]. Besides that, several recent trends are to integrate frequency control function with optimal economic dispatch [21], [22], or the decentralized active demand response [23], [24].

Most of the methods in TABLE I demonstrated their good control performance by simulations. Several theoretical results have been found for analyzing the stability or robust stability of the systems. However, such results are commonly based on centralized strategies under the assumption that the entire system can be effectively known as a whole. For instance, to design any *i*th controller, it is essential to know the details of not only other *N*-1 local subsystems but also the network structure. This assumption is not feasible for large-scale

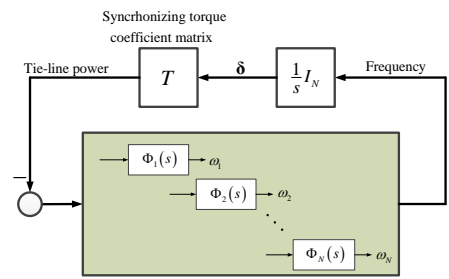

Fig. 1. A general expression of frequency control system.

systems of massive agents with various local dynamics.

When considering only frequency stability control, a large-scale power network can be treated as a multi-agent system of $N$ local agents interconnection via the synchronizing torque coefficient matrix $T$ of size $N \times N$ (Fig. 1). In general, $\Phi_i(s)$ is quite complex. It includes the transfer functions of local generator, turbine, governor, and the local controller. As proved by Tan *et al*, system stability is determined by the stability of the matrix $\Psi(s) = \det(I_N + diag\{\Phi_i(s)\}T/s)$ where $I_N$ is the unity matrix of size $N$ [15], [16]. It is very complex to analyze the stability of $\Psi(s)$ directly, especially when $N$ is a big number, and/or parameter uncertainties are introduced.

This paper is our very first step toward decentralized stabilization of frequency control system. We notice that the properties of matrix $T$ was not utilized effectively in almost previous works. In the frequency control problems, matrix $T$ can be considered as a symmetric positive semidefinite matrix. This motivates us to employ two existing theories: passivity and generalized frequency variable (GFV). This paper suggests that the former is suitable for frequency control with swing equation modelling of generators. On the other hand, the later method is applicable to LFC using area-control-error (ACE).

Passivity based control (PBC) was firstly proposed three decades ago as a solution to rigid robots [25]. It has been widely applied in various fields, especially networked control [26], [27], [28]. In recent years, PBC has been utilized for electric vehicle [29], optimal demand response [30], and stability analysis of power network with heterogeneous nonlinear bus dynamics [31]. In this paper, we utilize passivity approach to generalize the problem studied by Andreasson *et al* in [19], [20]. Based on the properties of matrix $T$, the swing dynamics is shown to be output strictly passive. Consequently, the frequency control system is stable if each local area is provided a passive controller. The passivity approach can be further applied to hierarchically decentralized frequency control systems with communication delays.

TABLE I
LITERATURE REVIEW OF RESEARCHES ON DECENTRALIZED FREQUENCY CONTROL FOR POWER SYSTEMS

| Ref. | Method | Comment |
|---|---|---|
| [7] | Robust control using Lyapunov theory | -Lyapunov theory is applied to the state space equation of the entire power systems.<br>-Simple simulation using a 2-area power-network. |
| [8] | Robust control using Lyapunov theory | -Lyapunov function is established for local subsystems. However, each local agent still needs to know the details of network parameters (i.e., the synchronizing torque coefficients).<br>-Simulation using a small-scale power network with 3 areas. |
| [9] | PID control design using maximum peak resonance specification | -No stability analysis.<br>-Simulation using a 2-area-power-network model. |
| [10] | Optimal AGC tuning by genetic algorithm | -Simple integral controller, and no stability analysis.<br>-Simulation using a simple power network with 2 areas. |
| [11] | Adjustment of AGC set-point by ANN | -No stability analysis.<br>-Control performance is demonstrated by simple simulation with 2-area-power-network. |
| [12] | Adaptive fuzzy gain scheduling for LFC | -No stability analysis.<br>-Demonstration of the control performance by 2-area-power-network model. |
| [13] | Structured singular value μ for LFC | -Although stability condition is given, it is still a centralized stabilization approach applied to LFC system model as a MIMO system.<br>-Simulation using a 4-area-power-network model. |
| [14] | MISO PID controller based on characteristic matrix eigenvalues | -Robust optimality of local MISO PID is presented, but the stability condition for the overall system is not discussed. |
| [15], 16] [17], [18] | Tuning of PID controller by IMC | -The conditions for stability and robust stability are based on the matrix of transfer function of the total system ($\Psi(s)$).<br>-Simulation using a 3-area and 4-area power-network models. |
| [19], [20] | Distributed PI control for LFC | -Stability condition is given by applying Routh-Hurwitz to the entire system's dynamics.<br>-Simulation using IEEE 30-bus test system, but the model is too simple: dynamics of turbines and governors are neglected, and the local agents are assumed to be homogeneous. |

The generalized frequency variable (GFV) was proposed in the last decade as one solution to analyze multi-agent systems [32], [33]. Its effectiveness was clarified by the applications of gene regulatory networks [34], platoon cars [35], and motion control of electric vehicles [36]. This paper applied GFV theory to the decentralized LFC problem. Let $H_i(s) = \Phi_i(s)/s$ be the transfer function of the local area. Let $\{H_n(s), \xi \in (0, 1)\}$ be a nominal model set to be shared among the local areas. If each local controller $C_i(s)$ is designed such that $H_i(s)$ follows the model set, i.e., $\Delta_i(s) = (H_i(s) - H_n(s))/H_n(s) \leq \xi$, then the system in Fig. 1 can be expressed as a homogeneous multi-agent system $\Sigma(H_n(s), -T)$ with perturbation $\Delta(s) = diag\{\Delta_i(s)\}$. Using this expression, we propose a triad of conditions that sufficiently guarantee frequency control system. The conditions are tested conveniently by a limited set of inequalities established from the GFV $\phi_n(s) = 1/H_n(s)$, the set of eigenvalues of matrix $-T$, and the volume $\xi$ of $\Delta(s)$.

The reminder of this paper is organized as follows. We discuss in Section 2 the works of Andreasson *et al* and Tan *et al* as two motivating examples. In Section 3, we investigate the application of passivity-based frequency control using swing dynamics. Then, Section 4 is to discuss the merits and current challenges of passivity approach. Section 5 shows that the decentralized LFC using area-control-error (ACE) can be effectively model by GFV. Then, the effectiveness of GFV theory is discussed in Section 6. Finally, the conclusions, remarks, and the future works are stated in Section 7.

## 2. Motivating examples

In this Section, we discuss two frequency control systems which have been commonly used for studying. Besides that, we examine the properties of physical interaction expressed by the synchronizing torque coefficient matrix.

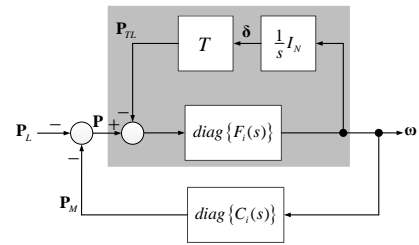

Fig. 2. Block diagram of frequency control system using swing equation.

### 2.1. Frequency control using swing equation

To study frequency control, many researchers modelled the power system as interconnected second-order system in which generator or aggregation of group of generators in an area is modelled by the swing equation. The decentralized frequency control system using swing equation is shown in Fig. 2. The local dynamics of frequency deviation is described by the inertial $M_i$ and the damping term $D_i$ as follows:

$$F_i(s) = \frac{1}{M_i s + D_i} \quad (1)$$

The physical interactions between the local areas are represented by the tie-line powers. To model the tie-line power, the synchronizing torque coefficient matrix is defined as

$$T = \begin{bmatrix} \sum_{j\neq 1} t_{1j} & -t_{12} & \cdots & -t_{1N} \\ -t_{21} & \sum_{j\neq 2} t_{2j} & \cdots & -t_{2N} \\ \vdots & \vdots & \ddots & \vdots \\ -t_{N1} & -t_{N2} & \cdots & \sum_{j\neq N} t_{Nj} \end{bmatrix} \quad (2)$$

where the synchronizing torque coefficient between area *i* and area *j*, namely $t_{ij}$, can be calculated from the voltages of aggregating buses *i* and *j* of areas *i* and *j*, and the susceptance of line (*i*, *j*) [1]. For every $i \neq j$, we have $0 \leq t_{ij} = t_{ji}$.

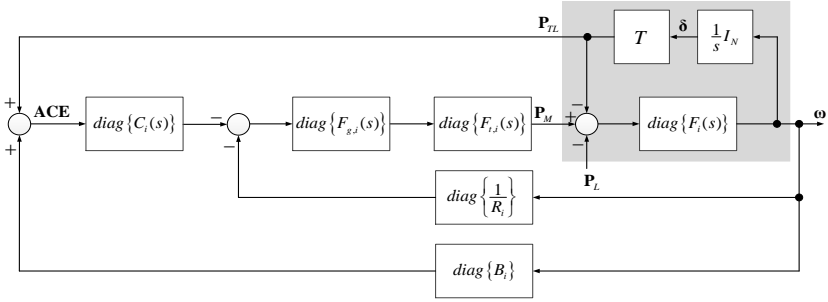
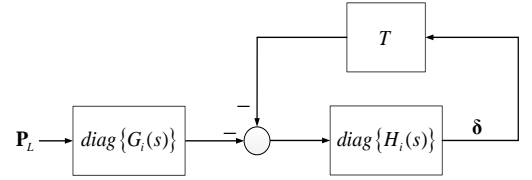

Fig. 3. Block diagram of decentralized load frequency control system using area control error.

Fig. 4. Equivalent expression of LFC system in Fig. 3.

TABLE II
NOMENCLATURE

| Symbol | Meaning |
|---|---|
| **Index** | |
| $i$ | Index of the local area ($i$ from 1 to $N$) |
| **Physical parameter** | |
| $M_i$ | Inertia |
| $D_i$ | Damping |
| $R_i$ | Droop characteristics |
| $B_i$ | Frequency bias setting |
| $t_{ij}$ | Synchronizing torque coefficient between $i$th and $j$th areas |
| $T$ | Synchronizing torque coefficient matrix |
| **Variable** | |
| $\omega_i$ | Frequency deviation |
| $\delta_i$ | Phase angle deviation |
| $ACE_i$ | Area control error |
| $P_{tl,i}$ | Tie-line power deviation |
| $P_{l,i}$ | Load variation |
| $P_{m,i}$ | Change of generation |
| **Vector of variables** | |
| $\boldsymbol{\omega}$ | $\boldsymbol{\omega} = [\omega_1 \ \omega_2 \ \ldots \ \omega_N]^T$ |
| $\boldsymbol{\delta}$ | $\boldsymbol{\delta} = [\delta_1 \ \delta_2 \ \ldots \ \delta_N]^T$ |
| **ACE** | $\mathbf{ACE} = [ACE_1 \ ACE_2 \ \ldots \ ACE_N]^T$ |
| $\mathbf{P}_{TL}$ | $\mathbf{P}_{TL} = [P_{tl,1} \ P_{tl,2} \ \ldots \ P_{tl,N}]^T$ |
| $\mathbf{P}_L$ | $\mathbf{P}_L = [P_{l,1} \ P_{l,2} \ \ldots \ P_{l,N}]^T$ |
| $\mathbf{P}_M$ | $\mathbf{P}_M = [P_{m,1} \ P_{m,2} \ \ldots \ P_{m,N}]^T$ |

We provide for each local area a controller $C_i(s)$. To stabilize the system in Fig. 2, a traditional way is to place the stable poles to the matrix of transfer function from to $\mathbf{P}_L$ to $\boldsymbol{\omega}$. This design scheme is not feasible due to the large size the power systems. In [19], Andreasson *et al* consider a special setting such that the local areas share a common PI controller

$$C_i(s) = \frac{as+b}{s} \quad \forall i \in [1,N] \tag{3}$$

By applying Routh-Hurwitz criterion, the frequency control system is shown to be stable if $a$ and $b$ are chosen as positive numbers. However, this result is not suitable for practical applications such that each local area has some degree of selfish. Since the local areas are heterogeneous, the assumption that they share the homogeneous controller is not so realistic. This discussion motivates us to generalize the stability condition stated in Ref. [19] for the frequency control system.

### 2.2. Load frequency control (LFC) based on ACE

The works of Andreasson *et al*, however, neglected the dynamics of some essential components as turbines and governors. Based on the classical two-area LFC system in the textbook of Kundur [1], we established the general LFC system in Fig. 3. This strategy utilizes the area control error defined as

$$ACE_i = P_{tl,i} + B_i \omega_i \tag{4}$$

The transfer functions of the turbine $F_{t,i}(s)$ and governor $F_{g,i}(s)$ depend on the type of generation, such as thermal generation and hydraulic generation. We neglect to present the expression of $F_{t,i}(s)$ and $F_{g,i}(s)$ which can be founded in [1] and [15] ~ [18]. The system in Fig. 3 can be equivalently expressed as the system in Fig. 4 where

$$H_i(s) = \frac{1}{s}\Phi_i(s) = \frac{1}{s} \frac{\left(1+C_i(s)F_{g,i}(s)F_{t,i}(s)\right)F_i(s)}{1+\left(B_i C_i(s)+\frac{1}{R_i}\right)F_{g,i}(s)F_{t,i}(s)F_i(s)} \tag{5}$$

$$G_i(s) = \frac{1}{1+C_i(s)F_{g,i}(s)F_{t,i}(s)} \tag{6}$$

Following the works of Tan *et al* [15]~[18], we have to verified by the stability of $\det(I_N + diag\{H_i(s)\}T)$. Unluckily, this calculation is feasible only for small-scale power systems of several interconnected areas. This motivates us to reduce the complexity of design and analysis of large-scale LFC systems.

### 2.3. Notice on physical interaction matrix

In this paper, we are interested in the following notice.

**Notice 1**: Power systems can be considered lossless in frequency control studies. Therefore, the synchronizing torque coefficient matrix can be treated as a symmetric positive semidefinite matrix with at least one zero eigenvalue. Moreover, there exist the unitary matrix that diagonalizes the synchronizing torque coefficient matrix $T$.

By elementary transformation, $T$ is shown to have at least one zero-eigenvalue. Define the nonzero vector $\mathbf{z} = [z_1 \ z_2 \ \ldots \ z_N]^T$. We have $\mathbf{z}^T T \mathbf{z} = 0.5 \sum_{i \neq j} t_{ij}(z_i - z_j)^2 \geq 0$ which clarifies that $T$ is positive semidefinite.

## 3. Passivity approach

### 3.1. Introduction to passivity theory

Let us consider a system given by a state space equation

$$\begin{cases} \dot{x} = f(x,u) \\ y = h(x,u) \end{cases} \tag{7}$$

where the input vector $u \in \mathbb{R}^p$, output vector $y \in \mathbb{R}^p$, and the state vector $x \in \mathbb{R}^n$.

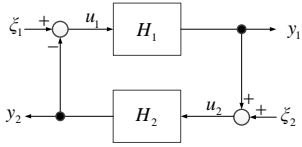

Fig. 5. Standard feedback connection of two subsystems.

**Definition 2** [37]: The system (7) is said to be passive from input *u* to output *y* if there exists a positive semidefinite function $S: \mathbb{R}^n \rightarrow \mathbb{R}_+$, called *storage function*, such that

$$\dot{S} = \frac{\partial S}{\partial x}\dot{x} \leq y^T u$$

holds for all $x \in \mathbb{R}^n$ and $u \in \mathbb{R}^p$.

In addition, (7) is said to be input strictly passive (ISP) if

$$\dot{S} \leq y^T u - \delta_u \|u\|^2, \quad \forall x \in \mathbb{R}^n, \forall u \in \mathbb{R}^p$$

for some scalars $\delta_u > 0$.

Finally, (7) is said to be output strictly passive (OSP) if

$$\dot{S} \leq y^T u - \delta_y \|y\|^2, \quad \forall x \in \mathbb{R}^n, \forall u \in \mathbb{R}^p$$

for some scalars $\delta_y > 0$.

In Fig. 5, let *H* be the feedback connection of two subsystems $H_1$ and $H_2$. The fundamental passivity theorems are stated as follows [36].

**Theorem 3-a**: If $H_1$ and $H_2$ are passive, then *H* is also passive with the input $(\xi_1, \xi_2)$ and the output $(y_1, y_2)$.

**Theorem 3-b**: If both subsystems $H_1$ and $H_2$ are OSP, then the closed loop system *H* with input $(\xi_1, \xi_2)$ and output $(y_1, y_2)$ has finite $L_2$-gain. When $\xi_2 = 0$, the closed loop system *H* with input $\xi_1$ and output $y_1$ has finite $L_2$-gain if $H_1$ is passive and $H_2$ is ISP, or $H_1$ is OSP and $H_2$ is passive.

### 3.2. Passivity analysis of swing dynamics

The frequency control system in Fig. 2 is the feedback connection of two subsystems: (i) The physical system in the shaded block, with the input $\mathbf{P} = -\mathbf{P}_M - \mathbf{P}_L$, and the output $\boldsymbol{\omega}$. (ii) The decentralized control system $C(s) = diag\{C_i(s)\}$ with the input $\boldsymbol{\omega}$, and the output $\mathbf{P}_M$.

**Proposition 4**: The physical system in Fig. 2 is OSP.

**Proof**: The physical system dynamics can be expressed as

$$M\dot{\boldsymbol{\omega}} + D\boldsymbol{\omega} + T\boldsymbol{\delta} = \mathbf{P} \tag{8}$$

where $M = diag\{M_i\}$ and $D = diag\{D_i\}$. As the damping terms are positive, matrix *M* is shown to positive definite. With respect to the **Notice 1**, we can define the storage function

$$S = \frac{1}{2}\boldsymbol{\omega}^T M \boldsymbol{\omega} + \frac{1}{2}\boldsymbol{\delta}^T T \boldsymbol{\delta} \tag{9}$$

From (8) and (9), we have

$$\boldsymbol{\omega}^T \mathbf{P} - \dot{S} = \boldsymbol{\omega}^T\left(M\dot{\boldsymbol{\omega}} + D\boldsymbol{\omega} + T\boldsymbol{\delta}\right) - \boldsymbol{\omega}^T M\dot{\boldsymbol{\omega}} - \boldsymbol{\omega}^T T\boldsymbol{\delta} = \boldsymbol{\omega}^T D\boldsymbol{\omega} \tag{10}$$

Select a small positive number $\rho$ such that $0 < \rho < min\{D_i\}$. Consequently, it is transparent that

$$\dot{S} = \boldsymbol{\omega}^T \mathbf{P} - \boldsymbol{\omega}^T D\boldsymbol{\omega} < \boldsymbol{\omega}^T \mathbf{P} - \rho \boldsymbol{\omega}^T \boldsymbol{\omega} \tag{11}$$

holds true for all input vector **P** and output vector $\boldsymbol{\omega}$. By the **Definition 2**, the physical system is said to be OSP with the storage function (9). This completes the proof.

### 3.3. Example 1: Decentralized frequency control

From **Theorem 3-b** and **Proposition 4**, the system in Fig. 2 can be stabilized by many options of decentralized controller.

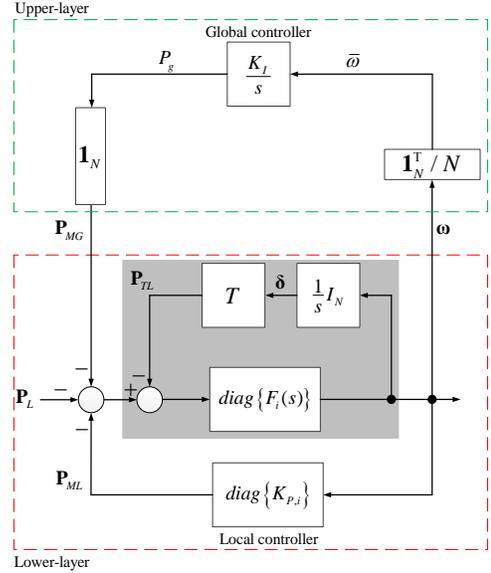

Fig. 6. Hierarchically decentralized frequency control system.

In general, we can select for each local area a passive control $C_i(s)$. For example, by the **Definition 2**, the class of PI controller with positive control gains is shown to be passive:

$$C_i(s) = \frac{K_{P,i}s + K_{I,i}}{s}, \quad K_{P,i} > 0, K_{I,i} > 0, \forall i \in [1, N] \tag{12}$$

### 3.4. Example 2: Hierarchical frequency control

This discussion is again motivated by the hierarchical frequency control system in [19] and [20]. As shown in Fig. 6, the system includes a global integral controller ($K_I/s$) and a bunch of local proportional controllers ($K_{P,i}$). The glocal controller is to control the average aggregation of frequency ($\bar{\omega}$), and distributes the additional control signal ($P_g$) to each local area. To obtain the average frequency deviation, we define $\mathbf{1}_N$ as an all-one-column-vector of size *N*. On the other hand, $\mathbf{P}_{ML}$ and $\mathbf{P}_{MG}$ are the vectors of generation command given by the local and global controllers.

**Proposition 5**: The hierarchically decentralized control system in Fig. 6 has finite $L_2$-gain with the selection of positive control gains $K_I$ and $K_{P,i}$ (*i* from 1 to *N*).

**Proof**: A sketch of the proof is presented as follows. Since $K_I$ is positive, the global controller is shown to be passive from $\bar{\omega}$ to $P_g$. The upper-layer system is a cascade connection of $\mathbf{1}_N^T / N$, the passive global controller, and $\mathbf{1}_N$. Consequently, it is also passive from $\boldsymbol{\omega}$ to $\mathbf{P}_{MG}$. This is a direct result of the Theorem 2.6 in [38]. Next, we consider the lower-layer system which is the feedback connection of the physical system and the local controllers. Since the local control gains $\{K_{P,i}\}$ are all positive, the lower-layer system is shown to be OSP from $-\mathbf{P}_{MG}$ to $\boldsymbol{\omega}$. Based on **Theorem 3-b**, the overall system in Fig. 6 has finite $L_2$-gain. This completes the proof.

### 3.5. Example 3: Hierarchical frequency control with unknown communication delay

Practically speaking, there exists the communication channels between the lower-layer and upper-layer. If the delay is not handled, the control performance and stability of the overall system would be degraded. We assume that there exists an unknown time delay $T_{d,i}^{l \to g}$ for sending the *i*th local frequency to the global controller. On the other hand, there exists an unknown time delay $T_{d,i}^{g \to l}$ for distributing the global control signal to the *i*th area. Applying the idea of scattering chain in [26], we implement two scatterings $S_G$ and $S_L$ as in Fig. 7. We assume that the global and local control gains are all positive. The vectors of size $N$ are defined as

$$\mathbf{v}_L = \begin{bmatrix} v_{L,1} & v_{L,2} & \cdots & v_{L,N} \end{bmatrix}^T, \mathbf{v}_R = \begin{bmatrix} v_{R,1} & v_{R,2} & \cdots & v_{R,N} \end{bmatrix}^T,$$
$$\mathbf{v}_G = \begin{bmatrix} v_{G,1} & v_{G,2} & \cdots & v_{G,N} \end{bmatrix}^T, \mathbf{u}_L = \begin{bmatrix} u_{L,1} & u_{L,2} & \cdots & u_{L,N} \end{bmatrix}^T,$$
$$\mathbf{u}_R = \begin{bmatrix} u_{R,1} & u_{R,2} & \cdots & u_{R,N} \end{bmatrix}^T, \mathbf{u}_G = \begin{bmatrix} u_{G,1} & u_{G,2} & \cdots & u_{G,N} \end{bmatrix}^T.$$

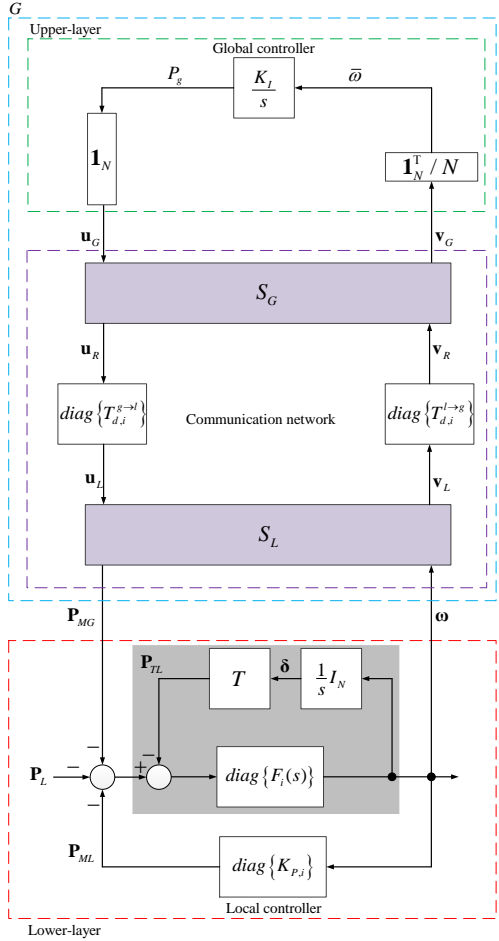

Fig. 7. Hierarchically decentralized frequency control system with communication delays.

The communication delays are expressed as
$v_{R,i}(t) = v_{L,i}(t - T_{d,i}^{l \to g})$, and $u_{L,i}(t) = u_{R,i}(t - T_{d,i}^{g \to l})$

The two scatterings have the diagonal structure, and they are implemented as $S_L = diag\{S_{L,i}\}$ and $S_G = diag\{S_{G,i}\}$ where

$$S_{L,i}: \begin{cases} u_{L,i} = \dfrac{P_{MG,i} - \alpha \omega_i}{\sqrt{2\alpha}} \\ v_{L,i} = \dfrac{P_{MG,i} + \alpha \omega_i}{\sqrt{2\alpha}} \end{cases}, S_{G,i}: \begin{cases} u_{R,i} = \dfrac{u_{G,i} - \alpha v_{G,i}}{\sqrt{2\alpha}} \\ v_{R,i} = \dfrac{u_{G,i} + \alpha v_{G,i}}{\sqrt{2\alpha}} \end{cases} \quad (13)$$

where $\alpha > 0$ is the parameter of the scatterings.

The energy balance for the communication network is now computed with the input vector and output vector respectively

$$\mathbf{u}_S = \begin{bmatrix} \boldsymbol{\omega}^T & -\mathbf{u}_G^T \end{bmatrix}^T, \mathbf{y}_S = \begin{bmatrix} \mathbf{P}_{MG}^T & \mathbf{v}_G^T \end{bmatrix}^T$$

$$\int_0^t \mathbf{y}_S^T \mathbf{u}_S d\tau = \int_0^t \left( \mathbf{P}_{MG}^T \boldsymbol{\omega} - \mathbf{v}_G^T \mathbf{u}_G \right) d\tau$$

$$= \frac{1}{2} \int_0^t \sum_{i=1}^N \left( u_{R,i}^2 - u_{L,i}^2 + v_{L,i}^2 - v_{R,i}^2 \right) d\tau \quad (14)$$

$$= \frac{1}{2} \sum_{i=1}^N \left( \int_{t-T_{d,i}^{g \to l}}^t u_{R,i}^2 d\tau \right) + \frac{1}{2} \sum_{i=1}^N \left( \int_{t-T_{d,i}^{l \to g}}^t v_{L,i}^2 d\tau \right) \geq 0 \quad \forall t$$

The inequality (14) shows that the communication network is passivated by the scatterings. Thus, the system $G$ in the blue dashed rectangular of Fig. 7 is passive, since it is a feedback connection of passive systems: the upper-layer system and the communication network. Consequently, the stability of the overall control system in Fig. 7 is guaranteed as a direct result of **Theorem 3-b**.

## 4. Discussion on passivity approach

### 4.1. Merit of passivity approach

From several applications presented in Section 3, passivity theory is shown to be an option for guaranteeing the stability of the frequency control system in a decentralized way. The complexity of the stability proof is considerably reduced thanks to the fundamental passivity theorems. In Example 1, each local area can select a controller (12) without the detailed understanding of the other areas and network structure given by matrix *T*. In Example 2, we can design in the upper-layer the global controller without any knowledge of the lower-layer.

Considering the existence of communication delays, Example 3 is an extension of Example 2. Thanks to the scattering transformation, we can design the upper-layer, the communication network, and the lower-layer independently. Motivated by the works of Lozano *et al*, in future study we will investigate the application of scattering chain to not only constant-delay but also time-varying delay systems [27].

We conducted the simulation of Example 3 to demonstrate the merit of passivity theory. The simulation parameters are listed in the TABLE V in **Appendix 1**. Unlike the homogeneous power system used in [19] and [20], we intentionally select a heterogeneous power system of ten local areas. The time-delays are unknown constants which might takes a value between 0.2195 [s] and 0.6585 [s]. The simulation is conducted from 0 to 100 seconds. At $t = 10$ [s], the changes of load are introduced to the local areas 3, 4, 5, 7, 8, and 10. Two test cases are conducted, and they share the same global integral control gain and local proportional control gains, as list in TABLE V. In Case 1, the system is not provided with scattering. In Case 2, the scattering is implemented with the parameter $\alpha = 0.6$.

Frequency deviation of Case 1 and Case 2 are described in Fig. 8(a) and Fig. 8(b), respectively. We notice that in Case 1, the system suffers frequency fluctuations. In Case 2, all frequency deviations smoothly converge to zero. Although the

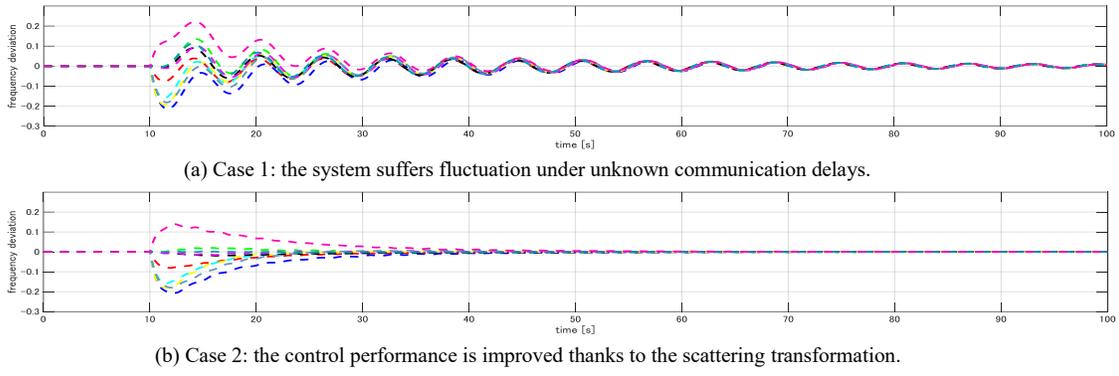

(a) Case 1: the system suffers fluctuation under unknown communication delays.

(b) Case 2: the control performance is improved thanks to the scattering transformation.

Fig. 8. Simulation results of frequency control using swing equation: Hierarchically decentralized control system with communication delays.

system in Case 2 is quite complex, its stability is guaranteed by a very convenient way.

### 4.2. Challenge of passivity approach

**How to address the existence of "actuators"?**

We reconsider the control system in Fig. 2. In actual power system, the control signal $\mathbf{P}_M$ is commonly realized by a bunch of local actuators with the transfer functions $\{G_{A,i}(s)\}$. Each $G_{A,i}(s)$ includes the transfer functions of the turbine and the governor. Consequently, the frequency control system in Fig. 2 becomes the feedback connection of the physical system in the shaded block and the equivalent controller $\tilde{C}(s) = diag\{\tilde{C}_i(s)\}$ where $\tilde{C}_i(s) = C_i(s)G_{A,i}(s)$. Even if $C_i(s)$ and $G_{S,i}(s)$ are all passive, the passivity of their series connection $\tilde{C}_i(s)$ is not guaranteed automatically. Theoretically speaking, if $\tilde{C}_i(s)$ is not passive, we can render its passivity by several ways, such as feedback passivation, feedforward passivation, or the combination of both [39]. However, the practical application of such methods requires further investigation. For instance, we need sensors to measure $\mathbf{P}_M$ for feedback passivation. We also need special mechanism to realize feedforward passivation. The input of this mechanism is frequency deviation, and the output has the same physical meaning as $\mathbf{P}_M$.

**How to connect the deterministic and stochastic?**

In recent year, decentralized active demand response (DADR) for power networks has been an interesting research topic [23], [24], [40], [41]. However, DADR is usually realized by a huge number of electric devices such as refrigerators. Unlike the turbines and governors, DADR is essentially modelled by stochastic characteristics, as discussed in [23] and [24]. We can only manage the probability that a local device is turned ON or OFF. Therefore, we cannot directly apply the fundamental passivity theorems to frequency control system with DADR. It is required to develop the passivity theory to integrate deterministic dynamical system expressed by swing equation with the stochastic mechanism of DADR.

### 5. Stabilization of LFC system with GFV

In Section 3, passivity theory is applied to analyze the frequency dynamics modelled by swing equation. This Section examines the LFC systems which includes the dynamics of governors and turbines. We firstly represent the LFC system by

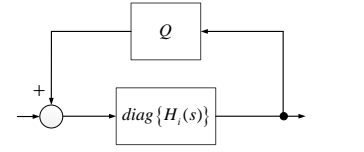

(a) Original system.

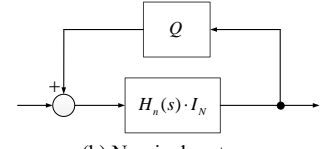

(b) Nominal system.

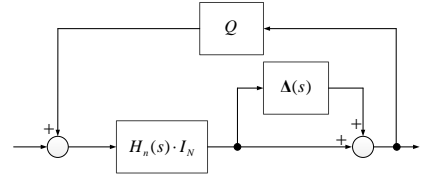

(c) Representation of the original system by perturbed system.

Fig. 9. Model of LFC system as multi-agent-system.

the nominal shared model set. Then, a triad of conditions that sufficiently guarantee the system stability is proposed. Finally, a procedure is established for designing LFC system.

### 5.1. Idea of shared model set

For simplicity, we define matrix $Q = -T$. As discussed in Sections 2, the stability of the LFC system in Fig. 3 is determined by the stability of the multi-agent-system shown in Fig. 9(a). The transfer function of a local agent, namely $H_i(s)$, is given by (5). Let $H_n(s)$ be the transfer function of a "nominal agent." The "nominal agent" is associated with a "model matching index" $\xi$ between 0 and 1. A model set is defined by a pair of $\{H_n(s), \xi\}$. The multiplicative error is calculated as

$$\Delta_i(s) = \frac{H_i(s) - H_n(s)}{H_n(s)} \tag{15}$$

If $\|\Delta_i(s)\|_\infty \leq \xi$, $H_i(s)$ is said to belong to $\{H_n(s), \xi\}$. If all the local agents belong to $\{H_n(s), \xi\}$, the system in Fig. 9(a) is represented by the system $\Sigma(H_n(s), Q, \Delta(s))$ in Fig. 9(c) where $\Delta(s) = diag\{\Delta_i(s)\}$, $\|\Delta(s)\|_\infty \leq \xi$. Here, $\Sigma(H_n(s), Q, \Delta(s))$ is obtained by introducing the perturbation $\Delta(s)$ of volume $\xi$ to the nominal homogeneous system $\Sigma(H_n(s), Q)$ in Fig. 9(b).

As explained in Section 1, it is very hard to guarantee the stability of the original multi-agent-system by examining the

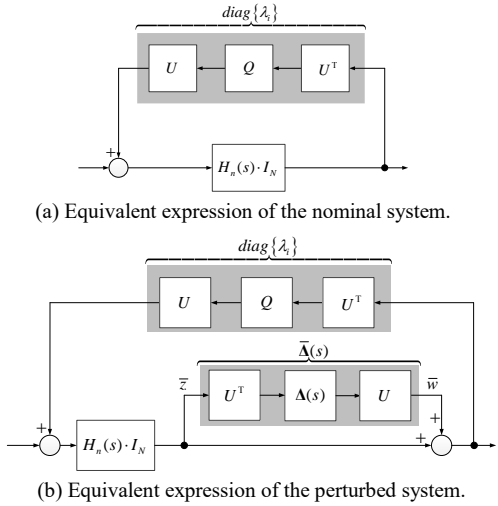

(a) Equivalent expression of the nominal system.

(b) Equivalent expression of the perturbed system.

Fig. 10. Equivalent expressions for the proposed tests.

matrix of transfer function $\det(I_N - diag\{H_i(s)\}Q)$. However, if $\Sigma(H_n(s), Q)$ is stable and $\Sigma(H_n(s), Q, \Delta(s))$ is robustly stable, we can say that the original multi-agent-system is stable (sufficiently). Therefore, it is essential to test the stability and robust-stability of the systems in Figs. 9(b) and (c), respectively. Applying GFV theory with respect to **Notice 1**, in the following sub-section, we will show that the complexity of such tests can be reduced considerably.

*5.2. Stability test and robust stability test*

Test (i): Stability test of $\Sigma(H_n(s), Q)$

From **Notice 1**, there exists a unitary matrix $U$ that diagonalizes matrix $Q$: $UQU^T = diag\{\lambda_i\}$ with $UU^T = U^TU = I_N$. Consequently, the nominal homogenous system $\Sigma(H_n(s), Q)$ is equivalent to the system in Fig. 10(a). Thus, $\Sigma(H_n(s), Q)$ is stable if and only if $H_n(s)\lambda_i \neq 1, \forall \lambda_i \in \sigma(Q), \forall s \in \mathbb{C}_+$ where $\sigma(Q)$ is the set of eigenvalues of matrix $Q$. Define the generalized frequency variable (GFV) $\phi_n(s) = 1/H_n(s)$. Since $Q$ has at least one zero eigenvalue, the stability condition of $\Sigma(H_n(s), Q)$ is reduced to

$$\phi_n(s) - \lambda_i \neq 0, \ \forall \lambda_i \in \sigma_-(Q), \forall s \in \mathbb{C}_+ \tag{16}$$

where $\sigma_-(Q)$ is the set of negative eigenvalues of matrix $Q$. Following the GFV theory [33], the following statement is obtained.

**Theorem 6**: The nominal system $\Sigma(H_n(s), Q)$ is stable if all the non-zero eigenvalues of matrix $Q$ are placed in the stable domain $\Omega_+^c$ defined as $\Omega_+ := \phi_n(\mathbb{C}_+)$, and $\Omega_+^c := \mathbb{C} \setminus \Omega_+$.

The stable domain is characterized by a limited set of LMIs established from the GFV [33]. The complexity level of the stability test, therefore, is independent of the number of the local agents. If the local agents $\{H_i(s)\}$ are quite homogeneous, we can select any local agent as the nominal one. Then, it is almost surely that Test (i) can guarantee the stability of the LFC systems. If the differences between the local agents are unneglectable, the volume of $\Delta(s)$ should be addressed by the following test.

Test (ii): Robust stability test of $\Sigma(H_n(s), Q, \Delta(s))$

A standard way to examine the robust stability of the system in Fig. 9(c) is to directly apply μ-analysis and synthesis [42]. To reduce the complexity of this test, we transform the system in Fig. 9(c) to the equivalent expression in Fig. 10(b) where the transfer function from $\bar{w}$ to $\bar{z}$ is diagonalized by

$$\bar{H}_i(s) = \frac{\lambda_i H_n(s)}{1 - \lambda_i H_n(s)}$$

If $\|\Delta(s)\|_\infty \leq \xi$, it follows that $\|\bar{\Delta}(s)\|_\infty \leq \xi$ as $U$ is the unitary matrix. The following statement is obtained as a direct result of the small gain theorem and Theorem 3 in Ref. [43].

**Proposition 7**: We assume that $\Sigma(H_n(s), Q)$ is stable and $\|\Delta(s)\|_\infty \leq \xi$, then the perturbed system in Fig. 9(c) is robustly stable if and only if

$$\left\|\frac{\lambda_i H_n(s)}{1 - \lambda_i H_n(s)}\right\|_\infty \leq \frac{1}{\xi} \ \forall \lambda_i \in \sigma_-(Q) \tag{17}$$

*5.3. A procedure for designing LFC systems*

From the theoretical results in the previous sub-section, we consider the following design procedure for the decentralized LFC system.

**Step 1**: Select a nominal model set $\{H_n(s), \xi\}$ such that (16) and (17) holds true for all non-zero eigenvalues of matrix $Q$.

**Step 2**: Design for each local area a control $C_i(s)$ such that the transfer function of the local agent $H_i(s)$ belongs to the nominal model set

$$\|\Delta_i(s)\|_\infty = \left\|\frac{H_i(s) - H_n(s)}{H_n(s)}\right\|_\infty \leq \xi \tag{18}$$

**Remark 8**: The above procedure includes a triad of conditions. The stability condition and robust stability condition of the nominal model set is given by (16) and (17), respectively. Besides, the model matching condition is given by (18). The procedure is to provide additional tests for many decentralized LFC methods which have been proposed. Each local area can design its $C_i(s)$ by either IMC method, LQR method, or robust control method… together with the model matching condition. The procedure sufficiently guarantees the stability of the original multi-agent-system. In other words, if we can find $\{H_n(s), \xi\}$ and a bunch of $\{C_i(s)\}$ that satisfying (16) ~ (18), we can say that the overall LFC system is stable. However, we cannot conclude that the overall LFC system is unstable if (18) is not satisfied for some local areas. Considering the model matching index, the following trade-off exists.

Scenario 1 (*$\xi$ is increasing and closed to* 1): This gives more degree of freedom to the local area since the model matching condition is easier to be satisfied. However, this also means the robust stability condition becomes harder to be satisfied.

Scenario 2 (*$\xi$ is decreasing and closed to* 0): It is harder for the local area to satisfy the model matching condition. However, it is easier to satisfy the robust stability condition.

When applying the procedure, it is essential to select the model matching index such that the trade-off is compromised to a certain extent. We will investigate the selection of the model matching index in future study. In the following Section, we only demonstrate the procedure and three conditions by a design example.

## 6. Design example and discussion

### 6.1. Problem setting

In [15] ~ [18], the numerical simulations are conducted using the power systems of three or four local areas. However, their models are quite homogeneous. For instance, in the three-area power system model in [16], the difference between the maximum inertia max$\{M_i\}$ and the minimum inertia min$\{M_i\}$ is only 1.3 times. The difference between the maximum damping max$\{D_i\}$ and the minimum damping min$\{D_i\}$ is only 1.07 times. On the other hand, the synchronizing torque coefficients $t_{ij}$ are set to 0.5 for all $i$, $j$. This means the eigenvalues of matrix $T$ are placed at only two points, $(0, 0)$ and $(-1.5, 0)$. From a practical point of view, the above setting is not so realistic. Moreover, the PID controller in [15]~[18] is

$$C_i(s) = K_{P,i} + \frac{K_{I,i}}{s} + K_{D,i}s$$

which is an improper transfer function.

This paper considers a power system of ten local areas as shown in **Appendix 2**. In this model, max$\{M_i\}$ equals two times of min$\{M_i\}$, and max$\{D_i\}$ equals 1.5 times of min$\{D_i\}$. The synchronizing torque coefficients have a large diversity in their values. Similarly to the simulation setting in [15]~[18], we consider the situation such that all local areas are provided with non-rehear thermal generation. The transfer function of the governors and turbines are expressed as

$$F_{g,i}(s) = \frac{1}{\tau_{g,i}s+1}, \quad F_{t,i}(s) = \frac{1}{\tau_{t,i}s+1} \tag{19}$$

where $\tau_{g,i}$ and $\tau_{t,i}$ are the time constant of the governor and turbine, respectively. Each area is provided a PID controller with the proper transfer function such that it can be implemented in real-time system

$$C_i(s) = K_{P,i}s + \frac{K_{I,i}}{s} + \frac{K_{D,i}s}{\tau_{D,i}s+1} \tag{20}$$

where $\tau_{D,i}$ is a small time constant. For simplicity, it is set to 0.01 for all local areas. Substitute (19) and (20) into (5), the transfer function of each local agent is

$$H_i(s) = \frac{b_{4,i}s^4 + b_{3,i}s^3 + b_{2,i}s^2 + b_{1,i}s + b_{0,i}}{s^6 + a_{5,i}s^5 + a_{4,i}s^4 + a_{3,i}s^3 + a_{2,i}s^2 + a_{1,i}s + a_{0,i}} \tag{21}$$

where the set of $\{a_{k,i}; b_{k,i}\}$ is shown in the **Appendix 2**.

### 6.2. Design example

**Step 1**: Selection of the nominal model set

We suppose that the nominal model set is established by a "global coordinator". The global coordinator calculated the nominal agent $H_n(s)$ by the nominal generator $F_n(s)$, the nominal governor $F_{gn}(s)$, the nominal turbine $F_{tm}(s)$, the nominal controller $C_n(s)$, the nominal bias $B_n$, and the nominal drooping gain $R_n$. The nominal transfer functions are

$$F_n(s) = \frac{1}{M_n s + D_n}, \quad F_{gn}(s) = \frac{1}{\tau_{gn}s+1}, \quad F_{tm}(s) = \frac{1}{\tau_{tm}s+1},$$

$$C_n(s) = K_{Pn}s + \frac{K_{In}}{s} + \frac{K_{Dn}s}{\tau_{Dn}s+1} \text{ where } \tau_{Dn} = 0.01.$$

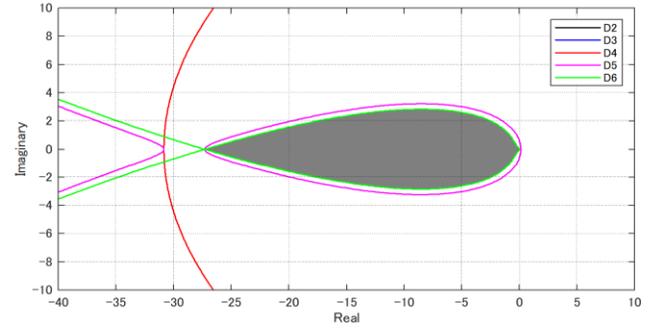

Fig. 11. Stable domain defined by $\phi_n(s) = 1/H_n(s)$ (shaded region).

A possible way to select the nominal inertia $M_n$ is that it should be between max$\{M_i\}$ and min$\{M_i\}$. By the similar way, we can select the other nominal parameters. In details, we select the set of nominal parameters as: $M_n = 3.43$, $D_n = 5.06$, $\tau_{gn} = 0.064$, $\tau_{tm} = 0.23$, $R_n = 2.50$, and $B_n = 0.417$. By placing the stable poles to the nominal transfer function

$$\frac{\left(1+C_n(s)F_{gn}(s)F_{tm}(s)\right)F_n(s)}{1+\left(B_nC_n(s)+\frac{1}{R_n}\right)F_{gn}(s)F_{tm}(s)F_n(s)}$$

the nominal PID gains are $K_{Pn} = 2.2820$, $K_{In} = 0.8244$, and $K_{Dn} = 0.3418$. The nominal agent is calculated as

$$\begin{aligned}H_n(s) &= \frac{1}{s}\frac{\left(1+C_n(s)F_{gn}(s)F_{tm}(s)\right)F_n(s)}{1+\left(B_nC_n(s)+\frac{1}{R_n}\right)F_{gn}(s)F_{tm}(s)F_n(s)}\\ &= \frac{b_{4n}s^4 + b_{3n}s^3 + b_{2n}s^2 + b_{1n}s + b_{0n}}{s^6 + a_{5n}s^5 + a_{4n}s^4 + a_{3n}s^3 + a_{2n}s^2 + a_{1n}s + a_{0n}}\\ &= \frac{0.2915s^4 + 34.95s^3 + 1309s^2 + 6413s + 1607}{s^6 + 121.3s^5 + 2230s^4 + 10020s^3 + 12500s^2 + 670s + 0}\end{aligned} \tag{22}$$

With respect to the trade-off described by **Remark 8**, we select the model matching index $\xi = 0.5$. Notice that this value is obtained by trial-and-error. The searching of the optimal $\xi$ is not considered in this paper.

Following Lemma 1 and Lemma 2 in [33], the stable domain is the shaded region in Fig. 11. It is determined by a set of LMIs $\{D_k(x, y) > 0\}$ where $k$ is from 1 to 6. The procedure to establish the LMIs are presented in **Appendix 2**. We have $D_1 = a_{5n} = 121.3 > 0$. Therefore, we only need to plot in Fig. 11 five curves $D_k = 0$ ($k$ from 2 to 6). The curves $D_{2,3} = 0$ do not appear since they exceed the range of the figure. The non-zero eigenvalues of matrix $Q$ are ranged between $-1.766$ and $-0.364$, which are certainly located in the shaded region. Based on **Theorem 6**, $\Sigma(H_n(s), Q)$ is stable.

Next, the nominal model set passes the robust stability test if

$$\kappa_i = \left\|\frac{\lambda_i H_n(s)}{1-\lambda_i H_n(s)}\right\|_\infty - \frac{1}{\xi} \leq 0 \quad \forall \lambda_i \in \sigma_-(Q) \tag{23}$$

The calculations of $\{\kappa_i\}$ are summarized in TABLE III. The results clarify that the perturbed system is robustly stable. This completes the **Step 1**.

**Step 2**: Local controller design

Now, the "global coordinator" distributes the nominal model set $\{H_n(s), \xi = 0.5\}$ to the local areas. This means the local controller can be designed with the degree of freedom given by $\xi$. By pole-placement to $\Phi_i(s) = sH_i(s)$, we design for each local area a PID controller. As summarized in TABLE IV, all the local controllers satisfy the model matching condition (18). This completes **Step 2** and the design procedure.

TABLE III
RESULTS OF ROBUST STABILITY TEST

| Non-zero eigenvalue $\lambda_i$ of matrix $Q$ | $\kappa_i = \left\|\dfrac{\lambda_i H_n(s)}{1-\lambda_i H_n(s)}\right\|_\infty - \dfrac{1}{\xi}$ | $\kappa_i \leq 0$ YES / NO |
|---|---|---|
| −1.766 | −0.7145 | YES |
| −1.432 | −0.7243 | YES |
| −1.288 | −0.7136 | YES |
| −1.059 | −0.7029 | YES |
| −0.926 | −1.6915 | YES |
| −0.826 | −0.6810 | YES |
| −0.680 | −0.6620 | YES |
| −0.600 | −0.6494 | YES |
| −0.364 | −0.6034 | YES |

TABLE IV
RESULTS OF MODEL MATCHING TEST

| Area No. | Local PID controller $C_i(s)$ | $\|\Delta_i(s)\|_\infty$ | $\|\Delta_i(s)\|_\infty \leq \xi$ YES / NO |
|---|---|---|---|
| 1 | $2.3800 + \dfrac{0.6716}{s} + \dfrac{0.4865}{0.01s+1}$ | 0.3979 | YES |
| 2 | $2.3250 + \dfrac{0.8381}{s} + \dfrac{0.3465}{0.01s+1}$ | 0.1286 | YES |
| 3 | $2.5880 + \dfrac{0.9681}{s} + \dfrac{0.3695}{0.01s+1}$ | 0.1561 | YES |
| 4 | $2.5880 + \dfrac{0.9681}{s} + \dfrac{0.3695}{0.01s+1}$ | 0.1549 | YES |
| 5 | $1.9850 + \dfrac{0.7115}{s} + \dfrac{0.2916}{0.01s+1}$ | 0.1201 | YES |
| 6 | $1.9930 + \dfrac{0.7775}{s} + \dfrac{0.2842}{0.01s+1}$ | 0.2411 | YES |
| 7 | $1.8950 + \dfrac{0.5669}{s} + \dfrac{0.3228}{0.01s+1}$ | 0.2086 | YES |
| 8 | $2.5590 + \dfrac{0.9264}{s} + \dfrac{0.3727}{0.01s+1}$ | 0.1709 | YES |
| 9 | $2.3270 + \dfrac{0.9104}{s} + \dfrac{0.3251}{0.01s+1}$ | 0.2252 | YES |
| 10 | $2.1360 + \dfrac{0.7066}{s} + \dfrac{0.3394}{0.01s+1}$ | 0.0939 | YES |

*6.3. Verification by simulations*

To evaluate the proposed design procedure, we conduct two simulation tests as follows. The simulation time is 100 seconds. At $t = 10$ seconds, small changes of load are introduced to the local areas 3, 4, 5, 7, 8, and 10.

Case 1-With power system modelled in **Appendix 2** (Fig. 12)
In this test, the simulation model is established using the parameters shown in **Appendix 2**. The simulation results, including the ACE and frequency variations are summarized in Figs. 12(a) and (b), respectively. They clarify that the overall LFC system is stable, and nice frequency control performance is attained.

Case 2-With power system modelled in **Appendix 3** (Fig. 13)
We notice that the actual power system is nonlinear, and the LFC system is obtained by linearization about an operating point. Therefore, it is essential that the control system has enough stability margin to tolerate the change of the linear model's parameters. To evaluate the controllers expressed in TABLE IV, the setting of Case 2 is as follows. We maintain the controller of Case 1 for conducting Case 2, in other words, $C_i^{\text{Case 2}}(s) = C_i^{\text{Case 1}}(s)$. However, we introduce large variations to the synchronizing torque coefficient matrix as:
$t_{ij}^{\text{Case 2}} = \alpha_{ij} t_{ij}^{\text{Case 1}} \quad \forall i \neq j$

where the variation gain $\alpha_{ij}$ is a number between 1.2 and 5.0. Next, we introduce the variations to the other parameters as follows:
$X_i^{\text{Case 2}} = \beta_i X_i^{\text{Case 1}}$ where $X_i \in \{M_i, D_i, \tau_{g,i}, \tau_{t,i}\}$

where the variation gain $\beta_i$ is a number between 0.92 and 1.06.

The model parameters of Case 2 are summarized in the **Appendix 3**. Again, the simulation results in Fig. 13 show that both frequency deviations and ACE deviations are smoothly suppressed. To explain the results of Case 2, we continue to use the nominal model set $\{H_n(s), \xi\}$ where $H_n(s)$ is given by (22) and $\xi = 0.5$. As presented in **Appendix 3**:

(i) The non-zero eigenvalues of the physical interaction matrix in Case 2 are placed between $-4.480$ and $-0.809$. From Fig. 11, such non-zero eigenvalues are certainly inside the stable domain given by $\phi_n(s) = 1/H_n(s)$. This means the nominal system $\Sigma(H_n(s), Q^{\text{Case 2}})$ is still stable.

(ii) The robust stability test shows that the perturbed system $\Sigma(H_n(s), Q^{\text{Case 2}}, \Delta(s))$ is robustly stable with $\|\Delta(s)\|_\infty \leq \xi = 0.5$.

(iii) Finally, the local controller $C_i(s)$ still satisfies the model matching condition.

From (i), (ii), and (iii), although LFC system suffers the large model parameter changes, it is still stable. Even if the eigenvalues of matrix $Q$ are increased to $-25$, system stability is still guaranteed.

*6.4. Discussion*

Applying GFV theory to LFC system, our goal is not to propose a new controller design method. We aim to provide three conditions (stability condition, robust stability condition, and model matching condition), and suggest a decentralized stabilization procedure for LFC of large-scale power systems. The procedure sufficiently guarantees system stability. Besides that, it can be used as an additional tool to analyze and support other LFC methods in literature. From the design example, each local area can be designed without understanding the details of the other controlled areas. The three conditions give an approach to figure out the system stability margin and measure the quality of the nominal shared model set. Towards the future steps, it is essential to examine the following issues.

*Firstly*, how should we select a nominal model effectively? In this paper, although the power system is heterogeneous, the differences between the local areas are not too large. We let the New England power system as an example [44]. This IEEE 39 bus system is a power model including ten local areas. The inertia of the area No. 1 is about nine times bigger than that of the areas No. 2 ~ No. 10. If $M_n$ is closed to the inertia values of the local area No. 2 ~ No. 10, then the model matching condition might become too strict for the area No. 1.

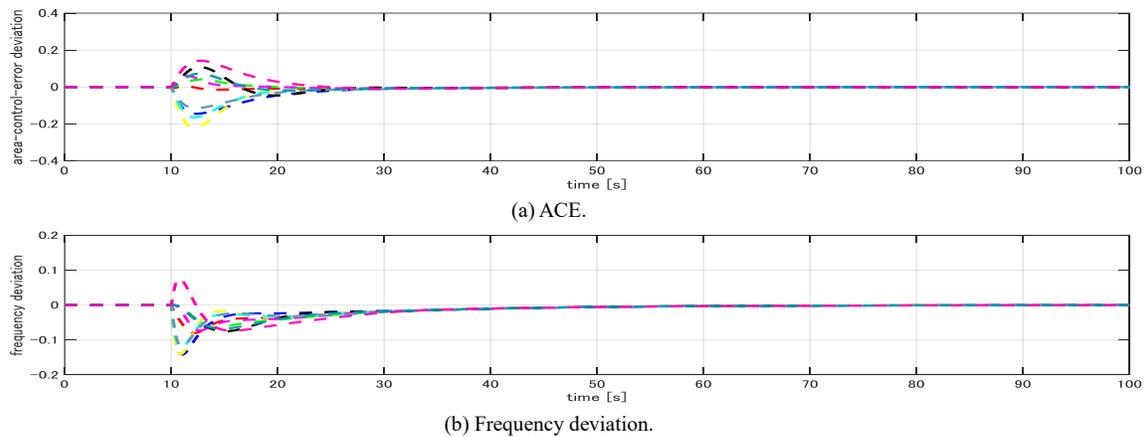

Fig. 12. Case 1: Simulation results of ACE based load frequency control – with parameter set in **Appendix 2**.

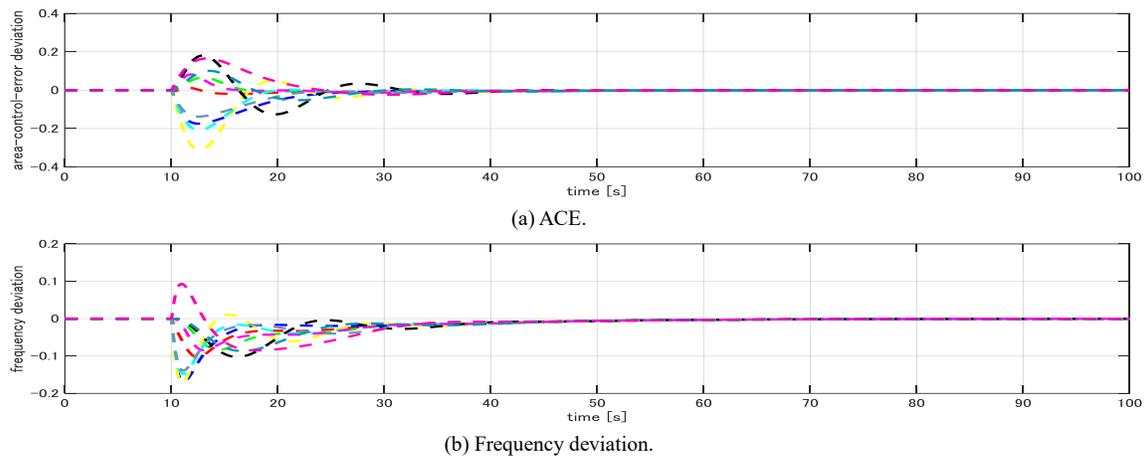

Fig. 13. Case 2: Simulation results of ACE based load frequency control – with parameter set in **Appendix 3**.

*Secondly*, how should we compromise the trade-off between the model matching condition and the robust stability condition. This issue has been briefly discussed by **Remark 8** of this paper. Given a nominal model, it is required to clarify the trade-off theoretically, or at least numerically.

*Thirdly*, from a practical point of view, it is essential to deal with $\mathcal{D}$-stability. The three conditions only guarantee that the poles of the LFC system are placed on the left-hand side of the complex plane. $\mathcal{D}$-stability is to guarantee not only stability, but also the damping ratio and convergence rate of the overall control system [33].

*Last but not least*, LFC considering measurement delay has been a topic of interest in recent years. This is due to the fact that the secondary control action usually suffers the measurement delay of the area-control-errors (ACEs). This delay might degrade the dynamics performance and cause instability. To deal with this issue, LMI theory is applied to design the single-area LFC system in [45], and interconnected-area LFC system in [46]. However, similarly to other works listed in TABLE I, the stability analysis in [46] is still a centralized approach. How to establish a nominal model and compensate the communication delays, especially time-varying delays, is a big challenge for LFC using GFV theory.

## 7. Conclusions

This paper presents a discussion on decentralized stabilization for frequency control of power networks. Our goal is to investigate the design procedures such that each area can be designed without the detailed understanding of the others. Motivated by the special properties of the physical interaction in the power networks, two theories are selected for our discussion. Considering the power networks modelled by swing equation, passivity theory is shown to be a candidate. On the other hand, GFV theory is shown to be a possible candidate for LFC system based on ACE. This paper is only a very first step towards decentralized stabilization of power networks. In future study, we are interested in the improvement of passivity approach such that it can be applied to complex power system models. Besides, how to select the nominal model and the model matching index, especially when the communication delay exists, is still a non-trivial question for GFV approach.

## Acknowledgement

This study is motivated from our work with Prof. Shinji Hara and Prof. Koji Tsumura (The University of Tokyo) on glocal (global/local) control of IWM electric vehicles. We would like to thank them for their kind advices.

TABLE V
PARAMETERS OF A TEN-AREA-POWER-NETWORK FOR FREQUENCY CONTROL USING SWING EQUATION

| Area No. | Inertia $M_i$ | Damping $D_i$ | Time-delay $T_{d,i}^{l \to g}$ | Time-delay $T_{d,i}^{g \to l}$ | Global control gain $K_I = 0.65$ |
|---|---|---|---|---|---|
| | | | | | Local control gains $\{K_{P,i}\}$ |
| 1 | 5.6 | 7.0 | 0.2634 [s] | 0.2195 [s] | 0.2800 |
| 2 | 3.2 | 4.4 | 0.5268 [s] | 0.4829 [s] | 0.2940 |
| 3 | 3.5 | 4.8 | 0.6146 [s] | 0.6585 [s] | 0.3080 |
| 4 | 3.0 | 5.1 | 0.5707 [s] | 0.6146 [s] | 0.3220 |
| 5 | 3.1 | 4.8 | 0.3073 [s] | 0.3512 [s] | 0.3360 |
| 6 | 2.8 | 5.2 | 0.3512 [s] | 0.3951 [s] | 0.3080 |
| 7 | 3.3 | 4.9 | 0.5927 [s] | 0.6146 [s] | 0.3220 |
| 8 | 3.5 | 4.6 | 0.4829 [s] | 0.5268 [s] | 0.2660 |
| 9 | 2.9 | 4.8 | 0.2634 [s] | 0.3073 [s] | 0.3500 |
| 10 | 3.4 | 5.5 | 0.5268 [s] | 0.5707 [s] | 0.3920 |

**Synchronizing torque coefficient matrix $T$**

$$T = \begin{bmatrix} 1.03 & -0.10 & -0.20 & -0.15 & -0.22 & -0.17 & -0.18 & 0 & 0 & -0.01 \\ -0.10 & 0.84 & -0.01 & -0.05 & -0.13 & -0.14 & -0.31 & -0.03 & -0.02 & -0.05 \\ -0.20 & -0.01 & 0.60 & 0 & -0.02 & -0.22 & -0.11 & -0.03 & -0.01 & 0 \\ -0.15 & -0.05 & 0 & 1.00 & 0 & -0.29 & -0.28 & -0.02 & -0.11 & -0.1 \\ -0.22 & -0.13 & -0.02 & 0 & 1.01 & -0.03 & -0.13 & -0.14 & -0.31 & -0.03 \\ -0.17 & -0.14 & -0.22 & -0.29 & -0.03 & 1.11 & 0 & 0 & -0.24 & -0.02 \\ -0.18 & -0.31 & -0.11 & -0.28 & -0.13 & 0 & 1.42 & -0.11 & -0.19 & -0.11 \\ 0 & -0.03 & -0.03 & -0.02 & -0.14 & 0 & -0.11 & 0.53 & 0 & -0.20 \\ 0 & -0.02 & -0.01 & -0.11 & -0.31 & -0.24 & -0.19 & 0 & 0.88 & 0 \\ -0.01 & -0.05 & 0 & -0.10 & -0.03 & -0.02 & -0.11 & -0.20 & 0 & 0.52 \end{bmatrix}$$

## Appendix 1

The simulation parameters of the ten-area power system model in Section 4 are summarized in TABLE V.

## Appendix 2

The power system model which was used for the design example and Case 1 of the simulation in Section 6, is expressed in TABLE VI. Based on this model, the transfer function $H_i(s)$ of the local agent is calculated with the set of $\{a_{k,i}; b_{k,i}\}$:

$$a_{5,i} = \frac{R_i \tau_{d,i} \left[ \tau_{g,i} \left( \tau_{t,i} D_i + M_i \right) + \tau_{t,i} M_i \right] + R_i \tau_{g,i} \tau_{t,i} M_i}{R_i \tau_{d,i} \tau_{g,i} \tau_{t,i} M_i},$$

$$a_{4,i} = \frac{R_i \left[ \tau_{d,i} \left( \tau_{g,i} D_i + \tau_{t,i} D_i + M_i \right) + \tau_{g,i} \left( \tau_{t,i} D_i + M_i \right) + \tau_{t,i} M_i \right]}{R_i \tau_{d,i} \tau_{g,i} \tau_{t,i} M_i},$$

$$a_{3,i} = \frac{R_i \left( \tau_{d,i} D_i + \tau_{g,i} D_i + \tau_{t,i} D_i + M_i \right) + B_i R_i \left( K_{D,i} + \tau_{d,i} K_{P,i} \right) + \tau_{d,i}}{R_i \tau_{d,i} \tau_{g,i} \tau_{t,i} M_i},$$

$$a_{2,i} = \frac{B_i R_i \left( K_{P,i} + \tau_{d,i} K_{I,i} \right) + 1 + R_i D_i}{R_i \tau_{d,i} \tau_{g,i} \tau_{t,i} M_i},$$

$$a_{1,i} = \frac{B_i R_i K_{I,i}}{R_i \tau_{d,i} \tau_{g,i} \tau_{t,i} M_i},$$

$$a_{0,i} = 0,$$

$$b_{4,i} = \frac{R_i \tau_{d,i} \tau_{g,i} \tau_{t,i}}{R_i \tau_{d,i} \tau_{g,i} \tau_{t,i} M_i},$$

$$b_{3,i} = \frac{R_i \left[ \tau_{d,i} \left( \tau_{g,i} + \tau_{t,i} \right) + \tau_{g,i} \tau_{t,i} \right]}{R_i \tau_{d,i} \tau_{g,i} \tau_{t,i} M_i},$$

$$b_{2,i} = \frac{R_i \left[ \tau_{d,i} + \tau_{g,i} + \tau_{t,i} + K_{D,i} + \tau_{d,i} K_{P,i} \right]}{R_i \tau_{d,i} \tau_{g,i} \tau_{t,i} M_i},$$

$$b_{1,i} = \frac{R_i \left[ K_{P,i} + \tau_{d,i} K_{I,i} + 1 \right]}{R_i \tau_{d,i} \tau_{g,i} \tau_{t,i} M_i},$$

$$b_{0,i} = \frac{R_i K_{I,i}}{R_i \tau_{d,i} \tau_{g,i} \tau_{t,i} M_i}.$$

With the nominal agent (22), the GFV is written as

$$\phi_n(s) = \frac{1}{H_n(s)} = \frac{s^6 + a_{5n}s^5 + a_{4n}s^4 + a_{3n}s^3 + a_{2n}s^2 + a_{1n}s + a_{0n}}{b_{4n}s^4 + b_{3n}s^3 + b_{2n}s^2 + b_{1n}s + b_{0n}}$$

Then, we establish the polynomial

$$pl = s^6 + a_{5n}s^5 + a_{4n}s^4 + a_{3n}s^3 + a_{2n}s^2 + a_{1n}s + a_{0n}$$
$$- (x + jy)(b_{4n}s^4 + b_{3n}s^3 + b_{2n}s^2 + b_{1n}s + b_{0n}) = s^6 + \sum_{i=1}^{6}(p_i + jq_i)s^{6-i}$$

Utilizing the Theorem 1 in [33], the set of $\{D_k\}$ is obtained as follows with the notice that $p_i = q_i = 0$ if $i > 6$.

$$\begin{cases} D_1 = \det |p_1| \\ D_k = \det \begin{bmatrix} F(\mathbf{p}_k) & -F(\mathbf{q}_k)R \\ UF(\mathbf{q}_k) & F(\mathbf{p}_{k-1}) \end{bmatrix} & (k = 2, ..., 6) \end{cases}$$

where

$$\mathbf{p}_k = \begin{bmatrix} 1 & p_1 & p_2 & \cdots & p_{2k-1} \end{bmatrix}^T$$

$$\mathbf{q}_k = \begin{bmatrix} 0 & 0 & q_1 & q_2 & \cdots & q_{2k-2} \end{bmatrix}^T$$

$$R = \begin{bmatrix} 0 \\ I \end{bmatrix} \in \mathbb{R}^{k \times (k-1)}, \quad U = \begin{bmatrix} I & 0 \end{bmatrix} \in \mathbb{R}^{(k-1) \times k}$$

$$F(x) = \begin{bmatrix} x_2 & x_4 & x_6 & \cdots & x_{2k} \\ x_1 & x_3 & x_5 & \ddots & \vdots \\ 0 & x_2 & x_4 & \ddots & x_{k+3} \\ \vdots & \ddots & \ddots & \ddots & x_{k+2} \\ 0 & \cdots & 0 & x_{k-1} & x_{k+1} \end{bmatrix}$$

TABLE VI
PARAMETERS OF A TEN-AREA-POWER-NETWORK FOR ACE BASED LOAD FREQUENCY CONTROL (DESIGN EXAMPLE AND SIMULATION-CASE 1, SECTION 6)

| Area No. | Inertia $M_i$ | Damping $D_i$ | Governor $\tau_{g,i}$ | Turbine $\tau_{t,i}$ | Droop $R_i$ | Bias $B_i$ | Synchronizing torque coefficient matrix $T$ |
|---|---|---|---|---|---|---|---|
| 1 | 5.6 | 7.0 | 0.068 | 0.26 | 2.52 | 0.41 | |
| 2 | 3.2 | 4.4 | 0.062 | 0.23 | 2.49 | 0.44 | |
| 3 | 3.5 | 4.8 | 0.060 | 0.22 | 2.51 | 0.43 | |
| 4 | 3.0 | 5.1 | 0.065 | 0.24 | 2.50 | 0.41 | |
| 5 | 3.1 | 4.8 | 0.066 | 0.22 | 2.53 | 0.38 | |
| 6 | 2.8 | 5.2 | 0.062 | 0.23 | 2.51 | 0.44 | |
| 7 | 3.3 | 4.9 | 0.070 | 0.25 | 2.48 | 0.41 | |
| 8 | 3.5 | 4.6 | 0.062 | 0.22 | 2.50 | 0.40 | |
| 9 | 2.9 | 4.8 | 0.058 | 0.20 | 2.51 | 0.42 | |
| 10 | 3.4 | 5.5 | 0.067 | 0.24 | 2.52 | 0.37 | |
| **Set of eigenvalues of matrix $Q = -T$** | | | | | | | $\sigma(Q)$ = {-1.766; -1.432; -; 1.288; -1.059; -0.926; -0.826; -0.680; -0.600; -0.364; 0} |

$$T = \begin{bmatrix} 1.03 & -0.10 & -0.20 & -0.15 & -0.22 & -0.17 & -0.18 & 0 & 0 & -0.01 \\ -0.10 & 0.84 & -0.01 & -0.05 & -0.13 & -0.14 & -0.31 & -0.03 & -0.02 & -0.05 \\ -0.20 & -0.01 & 0.60 & 0 & -0.02 & -0.22 & -0.11 & -0.03 & -0.01 & 0 \\ -0.15 & -0.05 & 0 & 1.00 & 0 & -0.29 & -0.28 & -0.02 & -0.11 & -0.1 \\ -0.22 & -0.13 & -0.02 & 0 & 1.01 & -0.03 & -0.13 & -0.14 & -0.31 & -0.03 \\ -0.17 & -0.14 & -0.22 & -0.29 & -0.03 & 1.11 & 0 & 0 & -0.24 & -0.02 \\ -0.18 & -0.31 & -0.11 & -0.28 & -0.13 & 0 & 1.42 & -0.11 & -0.19 & -0.11 \\ 0 & -0.03 & -0.03 & -0.02 & -0.14 & 0 & -0.11 & 0.53 & 0 & -0.20 \\ 0 & -0.02 & -0.01 & -0.11 & -0.31 & -0.24 & -0.19 & 0 & 0.88 & 0 \\ -0.01 & -0.05 & 0 & -0.10 & -0.03 & -0.02 & -0.11 & -0.20 & 0 & 0.52 \end{bmatrix}$$

TABLE VII
PARAMETERS OF A TEN-AREA-POWER-NETWORK FOR ACE BASED LOAD FREQUENCY CONTROL (SIMULATION-CASE 2, SECTION 6)

| Area No. | Inertia $M_i$ | Damping $D_i$ | Governor $\tau_{g,i}$ | Turbine $\tau_{t,i}$ | Droop $R_i$ | Bias $B_i$ | Synchronizing torque coefficient matrix $T$ |
|---|---|---|---|---|---|---|---|
| 1 | 5.21 | 6.65 | 0.072 | 0.239 | 2.52 | 0.41 | |
| 2 | 2.98 | 4.18 | 0.066 | 0.212 | 2.49 | 0.44 | |
| 3 | 3.26 | 4.56 | 0.064 | 0.202 | 2.51 | 0.43 | |
| 4 | 2.79 | 4.84 | 0.069 | 0.221 | 2.50 | 0.41 | |
| 5 | 2.88 | 4.56 | 0.070 | 0.202 | 2.53 | 0.38 | |
| 6 | 2.60 | 4.94 | 0.066 | 0.212 | 2.51 | 0.44 | |
| 7 | 3.07 | 4.65 | 0.074 | 0.230 | 2.48 | 0.41 | |
| 8 | 3.25 | 4.37 | 0.065 | 0.202 | 2.50 | 0.40 | |
| 9 | 2.69 | 4.56 | 0.064 | 0.202 | 2.51 | 0.42 | |
| 10 | 3.16 | 4.75 | 0.071 | 0.2208 | 2.52 | 0.37 | |
| **Set of eigenvalues of matrix $Q = -T$** | | | | | | | $\sigma(Q)$ = {-4.480; -3.391; -; 3.070; -2.593; -1.960; -1.703; -1.336; -1.212; -0.809; 0} |

$$T = \begin{bmatrix} 2.29 & -0.50 & -0.75 & -0.22 & -0.44 & -0.17 & -0.18 & 0 & 0 & -0.03 \\ -0.50 & 2.20 & -0.05 & -0.18 & -0.19 & -0.14 & -0.93 & -0.09 & -0.04 & -0.07 \\ -0.75 & -0.05 & 1.49 & 0 & -0.01 & -0.33 & -0.22 & -0.03 & -0.01 & 0 \\ -0.22 & -0.18 & 0 & 2.37 & 0 & -1.08 & -0.42 & -0.04 & -0.11 & -0.30 \\ -0.44 & -0.19 & -0.10 & 0 & 2.23 & -0.15 & -0.48 & -0.21 & -0.62 & -0.03 \\ -0.17 & -0.14 & -0.33 & -1.08 & -0.15 & 2.27 & 0 & 0 & -0.36 & -0.04 \\ -0.18 & -0.93 & -0.22 & -0.42 & -0.48 & 0 & 3.66 & -0.55 & -0.71 & -0.16 \\ 0 & -0.09 & -0.03 & -0.04 & -0.21 & 0 & -0.55 & 1.22 & 0 & -0.30 \\ 0 & -0.04 & -0.01 & -0.11 & -0.62 & -0.36 & -0.71 & 0 & 1.85 & 0 \\ -0.03 & -0.07 & 0 & -0.30 & -0.03 & -0.04 & -0.16 & -0.30 & 0 & 0.94 \end{bmatrix}$$

## Appendix 3

The power system model which was used for Case 2 of the simulation in Section 6, is expressed in TABLE VII. The results of the robust stability test and model matching test are summarized in TABLE VIII and TABLE IX, respectively. The results clarify that the LFC system is still stable in Case 2. Notice that we maintain the controller of Case 1 for conducting Case 2, in other words, $C_i^{\text{Case 2}}(s) = C_i^{\text{Case 1}}(s)$, for all index $i$ from 1 to 10.

TABLE VIII
RESULTS OF ROBUST STABILITY TEST (SIMULATION-CASE 2)

| Non-zero eigenvalue $\lambda_i$ of matrix $Q$ | $\kappa_i = \left\| \dfrac{\lambda_i H_n(s)}{1-\lambda_i H_n(s)} \right\|_\infty - \dfrac{1}{\xi}$ | $\kappa_i \le 0$ YES / NO |
|---|---|---|
| -4.480 | -0.2051 | YES |
| -3.391 | -0.4647 | YES |
| -3.070 | -0.5381 | YES |
| -2.593 | -0.6154 | YES |
| -1.960 | -0.7107 | YES |
| -1.703 | -0.7156 | YES |
| -1.336 | -0.7163 | YES |
| -1.212 | -0.7229 | YES |
| -0.809 | -0.6791 | YES |

TABLE IX
RESULTS OF MODEL MATCHING TEST (SIMULATION-CASE 2)

| Area No. | Local PID controller $C_i(s)$ | $\|\Delta_i(s)\|_\infty$ | $\|\Delta_i(s)\|_\infty \le \xi$ YES / NO |
|---|---|---|---|
| 1 | $2.3800 + \dfrac{0.6716}{s} + \dfrac{0.4865}{0.01s+1}$ | 0.3979 | YES |
| 2 | $2.3250 + \dfrac{0.8381}{s} + \dfrac{0.3465}{0.01s+1}$ | 0.1286 | YES |
| 3 | $2.5880 + \dfrac{0.9681}{s} + \dfrac{0.3695}{0.01s+1}$ | 0.1561 | YES |
| 4 | $2.5880 + \dfrac{0.9681}{s} + \dfrac{0.3695}{0.01s+1}$ | 0.1549 | YES |
| 5 | $1.9850 + \dfrac{0.7115}{s} + \dfrac{0.2916}{0.01s+1}$ | 0.1201 | YES |
| 6 | $1.9930 + \dfrac{0.7775}{s} + \dfrac{0.2842}{0.01s+1}$ | 0.2411 | YES |
| 7 | $1.8950 + \dfrac{0.5669}{s} + \dfrac{0.3228}{0.01s+1}$ | 0.2086 | YES |
| 8 | $2.5590 + \dfrac{0.9264}{s} + \dfrac{0.3727}{0.01s+1}$ | 0.1709 | YES |
| 9 | $2.3270 + \dfrac{0.9104}{s} + \dfrac{0.3251}{0.01s+1}$ | 0.2252 | YES |
| 10 | $2.1360 + \dfrac{0.7066}{s} + \dfrac{0.3394}{0.01s+1}$ | 0.0939 | YES |